\documentclass[onecolumn, notitlepage]{revtex4-1}
\usepackage[centering,hmargin=3.3cm,vmargin=2.5cm]{geometry}

\pdfoutput=1
\usepackage{amsmath}    
\usepackage{graphicx}   
\usepackage{verbatim}   
\usepackage{color}      
\usepackage{subfigure}  
\usepackage[hidelinks]{hyperref}   
\usepackage[percent]{overpic}
\usepackage{bm, upgreek}

\usepackage[overlay,absolute]{textpos}
\newcommand\PlaceText[3]{%
	\begin{textblock*}{10in}(#1,#2)
		#3
	\end{textblock*}
}%

\newcommand{\mum}{$\upmu$m }
\newcommand{\mumN}{$\upmu$m}

\begin{document}
	
\title{Mode-locked dysprosium fiber laser:\\picosecond pulse generation from 2.97 to 3.30~\mumN}
\author{R. I. Woodward$^*$, M. R. Majewski and S. D. Jackson}
\affiliation{MQ Photonics, School of Engineering, Macquarie University, New South Wales, Australia}
	
\date{\today}
	
\begin{abstract}
Mode-locked fiber laser technology to date has been limited to sub-3~\mum wavelengths, despite significant application-driven demand for compact picosecond and femtosecond pulse sources at longer wavelengths.
Erbium- and holmium-doped fluoride fiber lasers incorporating a saturable absorber are emerging as promising pulse sources for 2.7--2.9~\mumN, yet it remains a major challenge to extend this coverage.
Here, we propose a new approach using dysprosium-doped fiber with frequency shifted feedback (FSF).
Using a simple linear cavity with an acousto-optic tunable filter, we generate $\sim$33~ps pulses with up to 2.7~nJ energy and 330~nm tunability from 2.97 to 3.30~\mum ($\sim$3000--3400 cm$^\mathrm{- 1}$)---the first mode-locked fiber laser to cover this spectral region and the most broadly tunable pulsed fiber laser to date.
Numerical simulations show excellent agreement with experiments and also offer new insights into the underlying dynamics of FSF pulse generation.
This highlights the remarkable potential of both dysprosium as a gain material and FSF for versatile pulse generation, opening new opportunities for mid-IR laser development and practical applications outside the laboratory.

\vspace{0.3cm}

$^*$ robert.woodward@mq.edu.au
\end{abstract}
	
\maketitle

\PlaceText{25mm}{9mm}{APL Photonics, Accepted 22nd August 2018}
	
\section{Introduction}
Mode-locked fiber lasers in the near-infrared (0.8--2.5~\mumN) are an established technology which underpins a multitude of important applications in research, medicine and industry.
Building on this success, there is currently strong demand to extend the wavelengths of these compact, high-brightness pulse sources into the mid-infrared (mid-IR), to enable a range of new applications such as breath analysis, laser surgery and polymer processing, which exploit absorption resonances of important organic molecules and technical materials in this region~\cite{Ebrahim-Zadeh2008}.

The past decade has seen great progress using fluoride glass (e.g. ZBLAN) fiber doped with erbium (Er) and holmium (Ho) ions to generate pulses in the 2.7--2.9~\mum range~\cite{Zhu2017a}.
In terms of mode-locking mechanisms, \emph{real} saturable absorbers (SAs) such as semiconductor films~\cite{Hu2014,Tang2015,Wei2017b} and nanomaterials~\cite{Zhu2016a} have been employed to generate self-starting picosecond pulses, while pulse durations as short as 180~fs (with energies exceeding 7~nJ) have been achieved using nonlinear polarization evolution (NPE---an \emph{artificial} SA) in ring cavity designs~\cite{Duval2015,Antipov2016a}.
Mode-locked mid-IR fiber lasers to date, however, have only scratched the surface of the mid-IR region, failing to satisfy the need for flexible short-pulse sources beyond 3~\mumN.
It should be noted that alternative devices have also been proposed, including interband and quantum cascade lasers (achieving 100s ns pulses)~\cite{Canedy2014} and nonlinear frequency conversion of near-IR sources (e.g. offering nanojoule-level femtosecond pulses through Raman-shifted fiber sources~\cite{Tang2016} and microjoule few-cycle pulses using bulk optical parametric chirped pulse amplification, OPCPA~\cite{Elu2017}). However, the compact footprint, simplicity and flexibility of fiber lasers makes them the preferred option for many practical applications.

Recently, dysprosium (Dy) has been identified as an ideal dopant for next-generation mid-IR fiber lasers: it offers an extremely broad emission cross-section, which has enabled lasing from 2.8~\mum to 3.4~\mum (importantly, spanning absorption features of multiple functional groups) ~\cite{Majewski2018} and can be in-band pumped for high efficiency~\cite{Woodward2018_watt}.
Dysprosium thus offer exciting opportunities for long-wavelength ultrashort pulse generation, but until this work, these had yet to be explored.
It should be noted that Er:ZBLAN fibers offer an additional transition for 3.4--3.8~\mum emission, which is a distinct yet complementary wavelength range to dysprosium.
With this transition, pulsed outputs have been achieved through gain-switching~\cite{Jobin2018} and Q-switching~\cite{Henderson-Sapir2018,Qin2018}. Additionally, a recent study reported Er:ZBLAN mode-locking at 3.5~\mum using a black phosphorous saturable absorber~\cite{Qin2018}, although the important autocorrelation trace could not be measured in order to fully characterize the temporal output; thus, the shortest recorded pulse beyond 3~\mum from a fiber laser to date is on the order of tens of nanoseconds~\cite{Jobin2018}.

In this Article, we broaden the spectral coverage of mode-locked fiber laser technology, demonstrating self-starting picosecond pulse generation from a Dy:ZBLAN fiber laser and notably utilizing a relatively understudied pulse generation concept based on frequency shifted feedback (FSF).
Our novel design employs an acousto-optic tunable filter (AOTF) to realize FSF-based mode-locking, eliminating the need for an SA, in addition to providing electronically controlled broadband tunability.
Despite being a robust mode-locking technique, FSF has received relatively little attention compared to SA or active mode-locking, yet is ideal for novel spectral regions such as the mid-IR where conventional pulse generation approaches / materials may be incompatible.
This demonstration of picosecond pulse generation from 2.97 to 3.30~\mum is, to our knowledge, the first dysprosium mode-locked laser, the first mode-locked fiber source in this region, and the mostly widely tunable mode-locked fiber laser to date, creating new opportunities to exploit the full potential of the mid-IR.
Before describing our experiments, we begin by briefly considering the pulse generation mechanism in contrast to more conventional approaches.

\subsection{Frequency Shifted Feedback Lasers}

\begin{figure}[bt]
	\centering
	\includegraphics{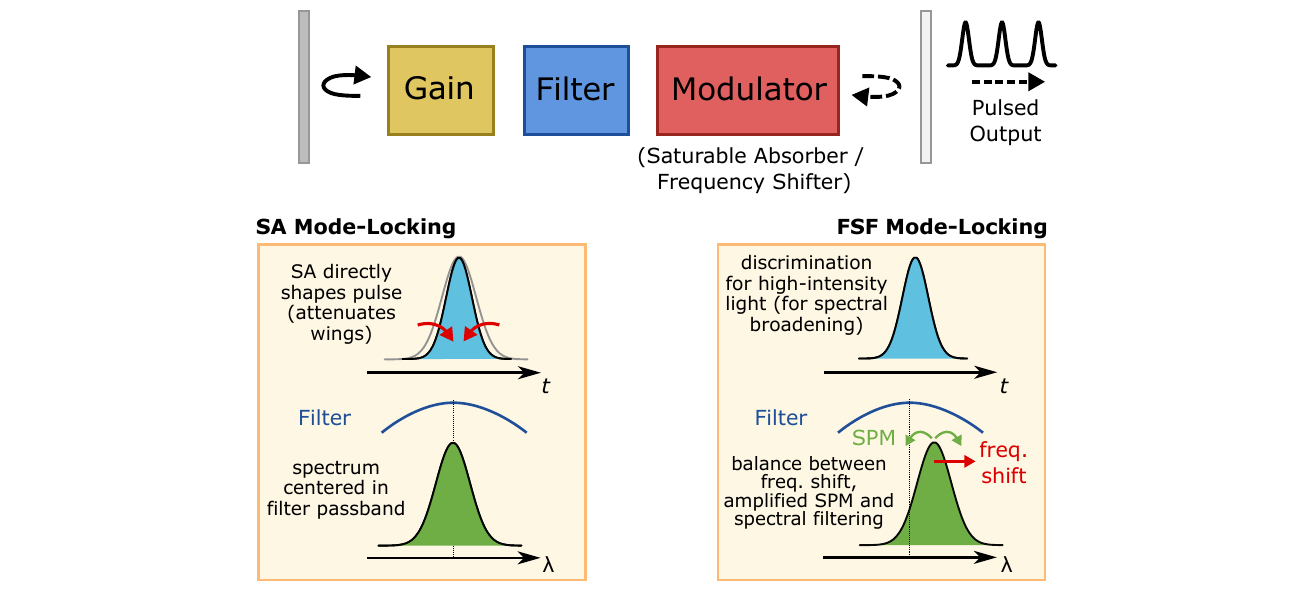}
	\caption{Illustration of a simple laser cavity comprising gain, spectral filtering and modulation (either a saturable absorber or frequency shifter), elucidating SA and FSF mode-locking mechanisms and their shaping effects in the temporal ($t$) and spectral ($\lambda$) domains.}
	\label{fig:concept}
\end{figure}

In a typical mode-locked laser, a loss modulator is included in the cavity to preferentially support a pulsed operating state instead of a continuous wave (CW) output.
This can be an actively driven modulator matched to the cavity round-trip frequency, which defines a short window of low loss within which a pulse can form, or an SA, where the nonlinear transmission of the device provides intensity discrimination, resulting in attenuation of low-intensity pulse wings and preferential amplification of high-intensity waveforms (Fig.~\ref{fig:concept}).
This leads to the generation of short pulses after many round trips, seeded by noise fluctuations.
The process can also be explained in the frequency domain by noting that modulation each round trip adds sidebands to each oscillating cavity mode.
As this modulation frequency is approximately equal to the cavity mode spacing, the sidebands injection-lock cavity modes to establish phase coherence across a broad bandwidth, corresponding to a pulse in the time domain.
If required, the exact emission wavelength can be defined by including a bandpass spectral filter.

In a frequency shifted feedback (FSF) laser, the modulator is replaced with a frequency shifter, yielding very different dynamics (Fig.~\ref{fig:concept}).
The frequency shifter monotonically shifts the wavelength of cavity light each round trip, which suppresses narrowband CW lasing since feedback is no longer resonant and continual wavelength shifting eventually pushes the signal outside the filter passband, leading to preferential amplification of incoherent spontaneous emission rather than the signal.
This has led to demonstrations of mode-less broadband sources where the light is effectively amplified spontaneous emission (ASE) and lacks coherence across the spectrum~\cite{Littler1991}---the temporal output is noise.

An FSF laser can generate short pulses, however, if nonlinearity is included in the cavity (e.g. nonlinear propagation in the gain fiber)~\cite{Cutler1992, Wu1993, Sabert1994}.
Specifically, intense light can cause spectral broadening through self-phase modulation (SPM), pushing energy from the center of the pulse spectrum against the direction of the frequency shift and enabling the signal to remain within the filter passband.
This is an intensity discriminating (i.e. self-amplitude modulation) effect, energetically favoring high-intensity short-pulse operation.
SPM-generated light is phase coherent and the round-trip frequency shift passes the phase of each spectral component to its neighbor; thus the FSF mechanism establishes a phase distribution across the emission bandwidth, i.e. supporting a coherent temporal pulse and resembling mode-locking.
It should be noted, however, that this cannot rigorously be described as `mode-locking' since FSF inhibits the build-up of longitudinal mode structure. Despite this, the laser output exhibits many characteristics that resemble typical mode-locking and in keeping with the existing literature~\cite{Sabert1994}, we adopt the term mode-locking to conveniently describe this operating state. 
Due to the steady-state energy balance between filtering and frequency shifting, the carrier frequency of the pulse can be slightly detuned from the peak of the filter passband.
The pulse repetition rate of an FSF laser is set by the cavity round trip time and the intracavity frequency shift does not need to be matched to the free spectral range of the cavity (frequency shifts down to a few kilohertz have enabled stable pulsation~\cite{Sabert1994}).

The origin of FSF pulse generation can be traced to pioneering 1960's work in phase-locking lasers using moving cavity mirrors (thereby introducing a Doppler frequency shift)~\cite{Henneberger1966}.
More recently, fiber laser work has exploited the FSF technique around 1~\mumN~\cite{Porta1998,Sabert1994,Heidt2007},  1.5~\mumN~\cite{Fontana1994,Sousa2000, Vazquez-Zuniga2014a} and 2~\mumN~\cite{Chen2016d}, typically generating pulses with tens of picosecond duration.
Ultrashort pulse generation has even been demonstrated by including additional pulse shortening mechanisms (e.g. NPE or soliton shaping)~\cite{Sousa2000,Kivisto2008,Noronen2016a,Heidt2007}, although compared to more conventional SA mode-locking, FSF laser dynamics and their prospects for high-power tunable sources are significantly understudied.
It is therefore a valuable and timely approach to consider FSF for the development of new turn-key mid-IR pulse sources.

\begin{figure}[b]
	\centering
	\includegraphics{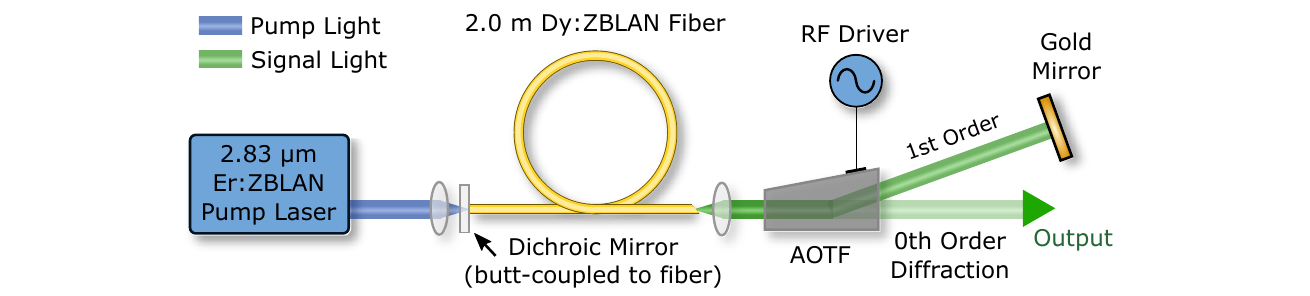}
	\caption{Schematic of the Dy-doped fiber FSF laser setup.}
	\label{fig:cavity}
\end{figure}

\section{Laser Design}
In this work, we employ a linear-cavity FSF laser design (Fig.~\ref{fig:cavity}), including 2.0~m of single-clad Dy-doped (2000 mol.\ ppm) ZBLAN fiber with 12.5~\mum core diameter and 0.16~NA.
A CW 2.83~\mum Er:ZBLAN fiber laser is employed as the pump source, corresponding to the Dy absorption cross-section peak.
A dichroic mirror (85\% transmissive at 2.8~\mum and  $>$95\% reflective for 3.0--3.5~\mumN, with reflectivity reducing from 95\% to 25\% from 3.0 to 2.9~\mumN) is butt-coupled to the input fiber facet and at the distal end, an external cavity (0.35~m length) is formed using a black diamond aspheric lens (after an 8$^\circ$ angle-cleaved fiber tip, to avoid back reflections) and an AOTF (Gooch \& Housego), where the first-order diffracted light is reflected back by a gold mirror to form the resonator.
Importantly, the AOTF is a traveling acoustic wave modulator, thus the diffracted light is Doppler shifted by a frequency equal to the RF sinusoidal drive signal (twice each round-trip due to the linear cavity design) that induces acoustic waves in the TeO$_2$ crystal.
The output is taken as the zeroth-order (undiffracted) light.
The AOTF has $\sim$1~dB insertion loss and a maximum diffraction efficiency of $\sim$75\% over the range 2 to 4~\mumN, where the center wavelength is set by the RF driving frequency and the 3-dB bandwidth is $\sim$5~nm at 3~\mumN.
The diffraction efficiency can be varied by changing the RF signal amplitude, enabling the cavity output coupling ratio to be varied in-situ.

\section{Numerical Simulations}
\label{sec:model}
We first consider numerical simulations to elucidate the FSF cavity dynamics and pulse build-up from noise.
Briefly, we employ a piecewise round-trip model where a field envelope (on a numerical grid in a co-moving reference frame at the envelope group velocity) is sequentially propagated through models for each cavity component over many round trips (described in detail in Ref.~\cite{Woodward2018_jo}). 
The AOTF is modeled as a Gaussian spectral filter followed by a frequency shift, which we set as 18.1~MHz for our experiment, corresponding to a 3.1~\mum filter central wavelength.
Care is taken to ensure the frequency shift is an integer multiple of the grid frequency spacing, to avoid numerical aliasing effects.
Fiber propagation is described by a generalized nonlinear Schr\"{o}dinger equation (GNLSE), including saturable gain, where the fiber dispersion and nonlinearity are computed from a step-index fiber eigenmode analysis~\cite{Snyder1983} (at 3.1~\mumN, group velocity dispersion $\beta_2=-148$~ps$^2$/km and nonlinear parameter $\gamma=0.22$~W$^{-1}$~km$^{-1}$).

\begin{figure}[b]
	\centering
	\includegraphics{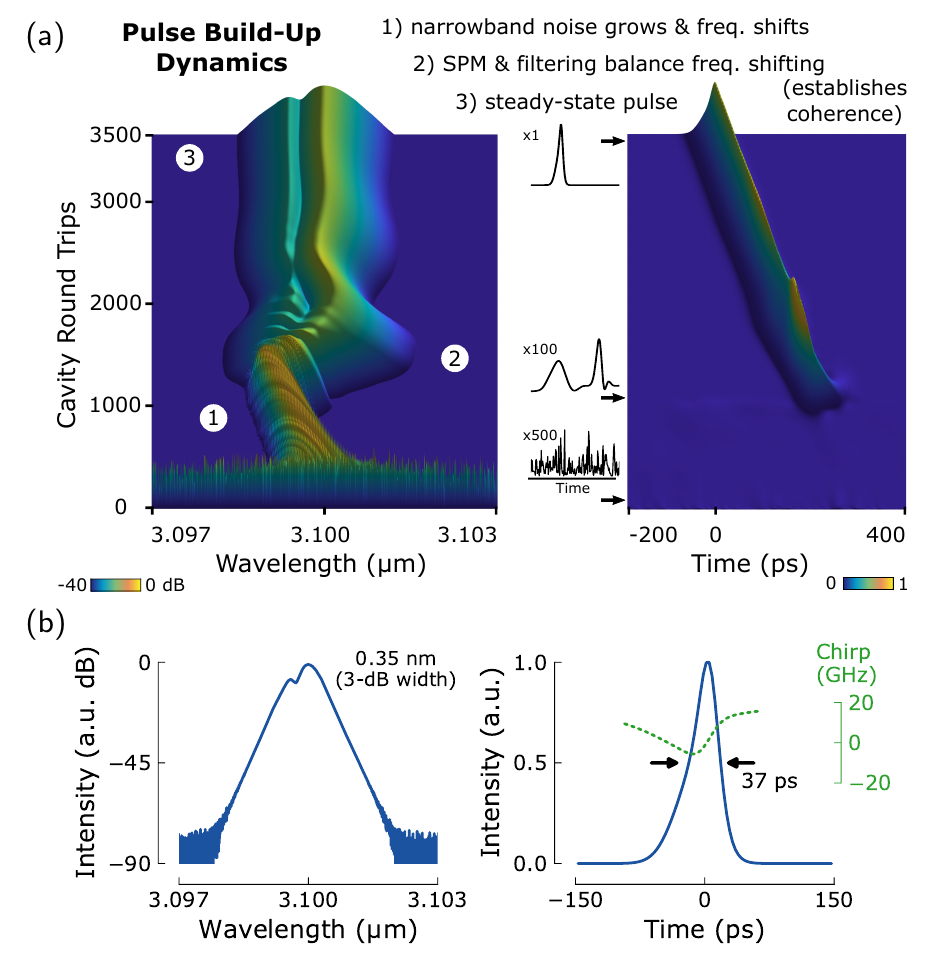}
	\caption{FSF laser simulation: (a) radiation build-up dynamics from noise; (b) steady-state pulse properties.}
	\label{fig:sim}
\end{figure}

The simulation is seeded with a shot noise field and over thousands of round trips converges to a steady-state 1.8~nJ pulsed solution [Fig.~\ref{fig:sim}(a)].
The nature of this evolution supports the mechanism of FSF pulse formation discussed earlier (and differs strongly from a typical mode-locked laser~\cite{Herink2016, Woodward_2016_pre}): as the field is initially amplified, it is shaped towards narrowband emission corresponding to temporal noise; for $\sim$100 to 1000 round-trips, the narrow emission steadily shifts in wavelength due to the AOTF frequency shift (a double-pass of 18.1~MHz shift at 3.1~\mum corresponds to 1.2~pm shift per round trip), although the temporal domain still primarily comprises noise; finally, as the signal wavelength is shifted away from the gain peak, nonlinearly seeded light under the peak is able to grow and while itself being frequency shifted (and inducing further nonlinear broadening)~\cite{Sabert1994}, causing the spectrum to broaden and settle into a steady state after approximately 2500 round trips.

Phase coherence across the bandwidth results in a pulse forming with 37~ps full width at half maximum (FWHM) duration and a slightly asymmetric shape (such asymmetry is typical for FSF fiber lasers~\cite{Sabert1994}).
The pulse is almost transform limited with only a small nonlinear chirp across it; there is thus limited prospects for pulse shortening through direct compression.
The spectral shape and notably, the `shoulder' on the short wavelength edge are noteworthy, as discussed later in comparison to our experiments.

Numerous iterations of the simulation from different random noise fields always converge to the same pulse output, confirming that this is a steady state for the system.
Investigation of the pulse at each position in the cavity during a single round trip reveal that there is negligible temporal / spectral breathing (i.e. duration and bandwidth are constant throughout the cavity).
Despite the fiber exhibiting Kerr nonlinearity and anomalous dispersion, soliton shaping is not involved in the dynamic and variation of the group velocity dispersion was found to have no effect on the pulse properties.
Such observations are in agreement with analytical studies confirming the steady-state FSF laser pulse is defined by a balance between gain and loss, rather than dispersion and nonlinearity~\cite{DeSterke1995}.

\section{Experimental Results}
To experimentally realize these promising modeling results, the FSF laser design in Fig.~\ref{fig:cavity} is constructed, with a nominal AOTF center wavelength of 3.1~\mum (RF drive frequency of 18.1~MHz and the RF power is set empirically to 1.2~W to maximize output power).
Lasing is observed at a threshold of 440~mW [Fig.~\ref{fig:exp}(a)].

\subsection*{Q-Switching}
Above threshold, the optical spectrum shows a narrow peak centered at 3.1~\mum with 0.28~nm FWHM [Fig.~\ref{fig:exp}(b)].
Temporally, the output is a periodic pulse train at 45~kHz repetition rate with 4.5~$\upmu$s duration pulses.
As the pump power is increased, the output power (measured after a filter to block any remaining pump light) increases up to 43 mW, the pulse repetition rate increases to 52~kHz and the pulse duration reduces to 2.5~$\upmu$s [Fig.~\ref{fig:exp}(c)], corresponding to a pulse energy of 0.83~$\upmu$J.
Such observations are characteristic of Q-switching.

While there is no Q-switch component implicitly included in the cavity, the observation of kilohertz pulse trains could be explained as sustained relaxation oscillations in the Dy:ZBLAN fiber.
Unlike a conventional laser where relaxation oscillations are damped once a steady-state inversion is reached, the FSF mechanism inhibits a CW steady-state by continually shifting the wavelength of intracavity light and thus, the instantaneous loss, inducing periodic self-sustaining Q-switching~\cite{Sabert1994,Bonnet1996,Perry1994}.
To verify the role of FSF in this pulsing dynamic, the cavity was adapted to resonate the undiffracted light from the AOTF with a 50\% mirror: in this case, light experiences no round-trip frequency shift and a pulsed output could not be observed.
Returning to the original FSF cavity, it was found that as the pump power was further increased, the laser transitioned from a Q-switched state to continuously mode-locked operation.

\subsection*{Mode-Locking}
\begin{figure}[bt]
	\centering
	\includegraphics{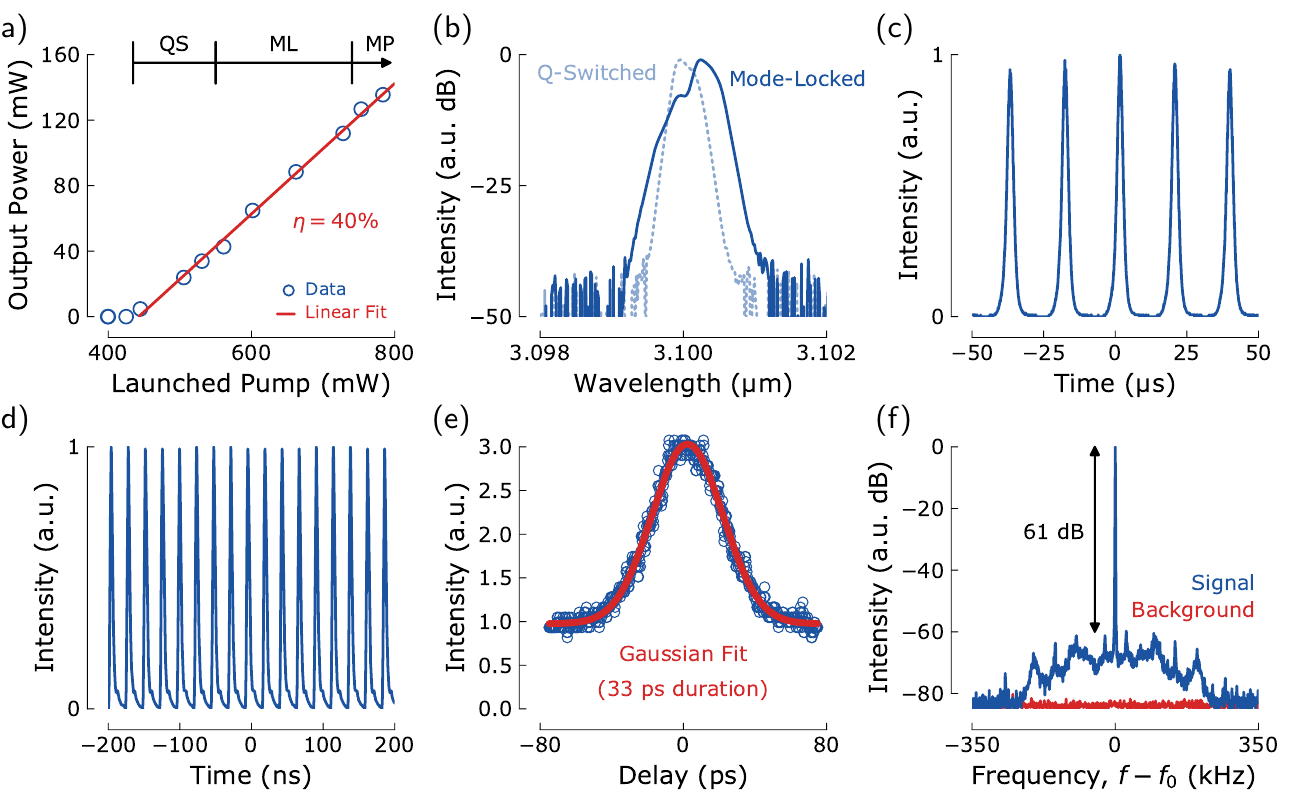}
	\caption{Experimental FSF laser performance: (a) power characteristic; (b) optical spectra. Q-switched (QS) operation at 43~mW output power: (c) oscilloscope trace. Mode-locked (ML) operation at 80~mW output: (d) oscilloscope trace; (e) collinear intensity autocorrelation trace; (f) RF spectrum of fundamental ($f_0=44.5$~MHz). Above 120~mW output, multi-pulsing mode-locking (MP) is observed.}
	\label{fig:exp}
\end{figure}

Above $\sim$550~mW pump power, self-starting mode-locking is observed, generating a stable pulse train with 44.5~MHz repetition rate [Fig.~\ref{fig:exp}(d)].
The spectrum changes significantly compared to the previous operating regime [Fig.~\ref{fig:exp}(b)], showing broadening to 0.33~nm FWHM, strong asymmetry and the appearance of an inflexion point / `shoulder'. 
A custom-built two-photon absorption autocorrelator is used to observe the output pulses, revealing a clean intensity autocorrelation trace [Fig.~\ref{fig:exp}(e)].
We are unable to precisely characterize the true pulse shape due to the symmetric nature of autocorrelation traces, although a Gaussian pulse shape can be well fitted to the measurement with a deconvolved pulse width of 33~ps.

To assess the stability of our laser, the output is observed with a photodetector connected to an RF spectrum analyzer: a train of harmonics are recorded, with a peak to pedestal value of $\Delta P=61$~dB at the fundamental frequency [Fig.~\ref{fig:exp}(f)]. 
From this measurement the pulse-to-pulse energy fluctuation can be approximated~\cite{Linde1986}: $\Delta E / E = \sqrt{\Delta P \Delta f / f_\mathrm{bw}}$ where $\Delta f\sim400$~kHz is the pedestal width and $f_\mathrm{bw}=300$~Hz is the resolution bandwidth.
We compute a low value of $\sim$3\% fluctuation, and note that the measured RF contrast is similar to mid-IR fiber lasers mode-locked with saturable absorbers~\cite{Zhu2016a,Tang2015,Wei2017b,Hu2014}, confirming the stability of our FSF laser as a picosecond pulse source, suitable for subsequent applications.

As the pump power is increased, the average output power of the pulse train increases until $\sim$750~mW incident power, when the pulse train becomes unstable, evolving into a multi-pulsing state as the power is further increased. 
Such instability is common for high-power pulsed oscillators, attributed to excess nonlinearity~\cite{Sabert1994}. 
The pulses appear approximately equally spaced on an oscilloscope, potentially indicating a harmonic mode-locking state with 89~MHz repetition rate, although we do not study this state further as our focus is primarily high energy pulse generation.

A linear relationship is observed between pump and output power, yielding an overall slope efficiency of 40\%.
For the fundamentally mode-locked pulse train, the maximum average power was $\sim$120~mW, corresponding to 2.7~nJ pulse energy.
Due to the birefringent nature of the AOTF, only one polarization component is resonated, leading to a strongly linearly polarized laser output, with a measured polarization extinction ratio (PER) of 24~dB. 

We note that there is excellent agreement between simulated and experimental mode-locked results, indicating that the model sufficiently captures the relevant dynamics.
It is particularly notable that the simulation correctly predicts the asymmetric spectral shape and `shoulder'.
Physically, this feature relates to the steady-state spectral energy distribution required to balance SPM-broadening, filtering and frequency shifting to maintain coherent mode-locked operation.
This phenomena has previously been observed in a near-IR FSF fiber laser and it was suggested that this balance was the cause of the asymmetry in the time domain too~\cite{Vazquez-Zuniga2014a}.
With a narrower bandpass filter, it may be possible to adjust the steady-state energy dynamic to suppress the spectral shoulder, improving the pulse quality, although in our experiment this is fixed by the AOTF.
While our autocorrelation measurement is unable to resolve asymmetry in the pulse, the estimated duration is still consistent with simulations.

\begin{figure}[bt]
	\centering
	\includegraphics{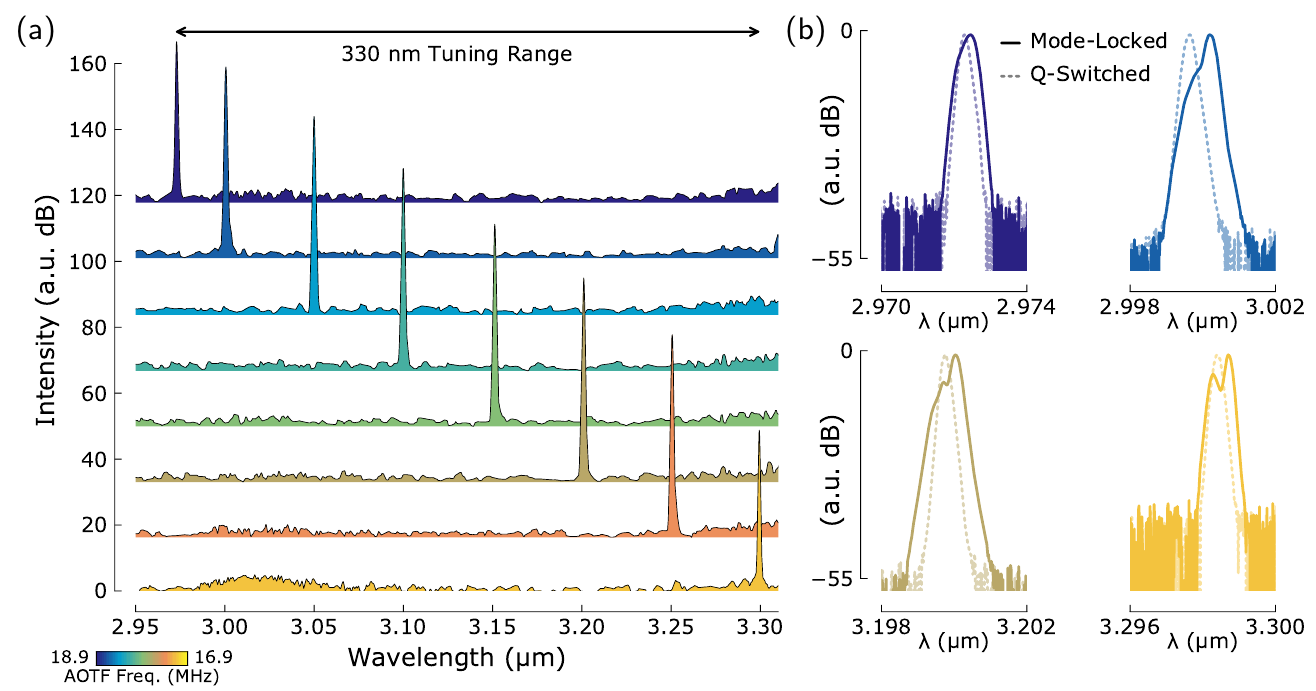}
	\caption{Characteristic spectra within the mode-locked FSF laser's 330 nm tuning range: (a) wide span (arbitrarily shifted in intensity for visual clarity); (b) close-up showing different spectral shapes between Q-switching and mode-locked operation.}
	\label{fig:tuning}
\end{figure}

\begin{figure}[bt]
	\centering
	\includegraphics{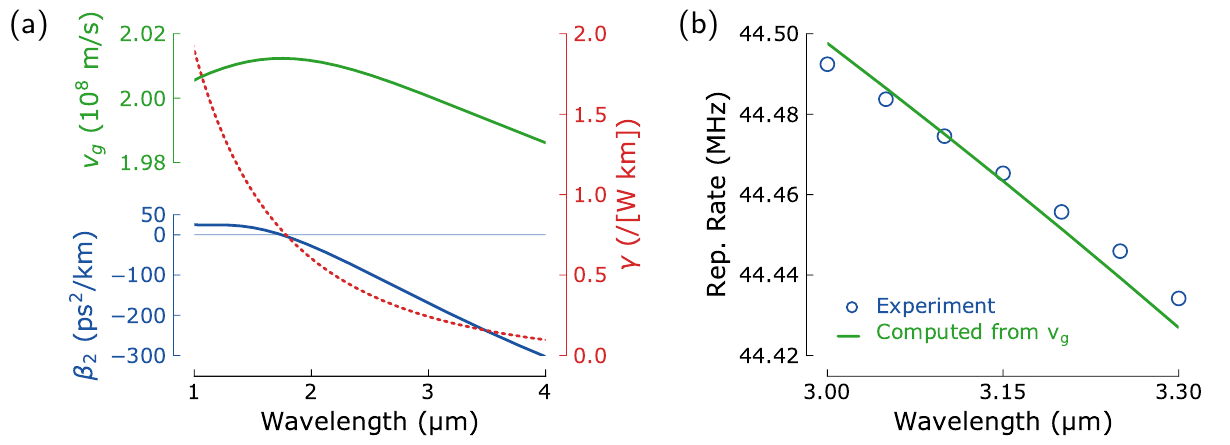}
	\caption{(a) Computed fiber group velocity $v_\mathrm{g}$, group velocity dispersion $\beta_2$, and nonlinear parameter $\gamma$. (b) Variation of mode-locked pulse repetition rate with laser wavelength.}
	\label{fig:reprate}
\end{figure}

Finally, we explore the tunability of our FSF laser by varying the AOTF RF frequency to shift the passband center frequency. 
Lasing is obtained from 2.95 to 3.30~\mumN, limited at the short wavelength edge by the reduced reflectivity of the dichroic cavity mirror (reflectivity $<50\%$ below 2.97~\mumN) and signal re-absorption (the Dy transition is quasi-three-level), and at the long wavelength edge by limited gain. 
Within this range, we are able to achieve stable mode-locking over the range 2.97--3.30~\mum (Fig.~\ref{fig:tuning}); at the spectral extents, we surmise that the gain is just sufficient for lasing, but not mode-locking.
When the laser is mode-locked, we note that continuous tunability by AOTF frequency adjustment is only possible over $\sim$40~nm, after which picosecond pulsation becomes unstable (e.g. multipulsing) or is lost completely. 
At different wavelengths, however, it is possible to adjust the system back into a self-starting pulsed state by varying the pump power (i.e. gain) or RF drive amplitude, which adjusts the output coupling ratio and thus, the cavity loss.
It is known that the AOTF RF power for optimum diffraction efficiency varies with wavelength, as does the fiber nonlinear parameter [Fig.~\ref{fig:reprate}(a)] and cavity loss due to lens chromatic aberration.
These effects change the net nonlinear phase shift and the gain/loss energy balance, which is required for stable operation as described earlier.
While we manually optimize these parameters to cover the wide tuning range here, we note that intelligent self-tuning algorithmic approaches are currently emerging for fiber lasers which could automatically self-optimize the performance as a future improvement~\cite{Andral2015, Woodward_scirep_2016}.

At each wavelength in the 330~nm tuning range, mode-locked operation is clearly distinguished from Q-switching at lower pump powers by spectral broadening and the characteristic asymmetric shape [Fig.~\ref{fig:tuning}(b)].
Additionally, autocorrelation measurements showed consistent pulse durations of 33$\pm$2~ps, with excellent stability ($>$60~dB RF contrast) in this region.
The pulse repetition rate is slightly decreased at longer wavelengths [Fig.~\ref{fig:reprate}(b)], due to longer cavity round trip times arising from lower group velocities (excellent agreement is noted when comparing experimental values to the predicted change from the computed fiber properties using an step-index fiber eigenmode analysis including a ZBLAN Sellmeier equation). 

\section{Discussion}

\subsection*{Opportunities with Dy:fiber}
This work showcases the excellent potential of the dysprosium ion for advancing mode-locked fiber laser technology by breaking  through the long-standing 3~\mum barrier.
As a result of the broad Dy emission cross section, mode-locking is achieved up to 3.3~\mumN, significantly improving upon the 2.7--2.9~\mum wavelengths achieved using Er and Ho ions.
The 330~nm tunability we observe significantly exceeds current best-in-class tunable mid-IR mode-locked fiber lasers (e.g.\ 34~nm tunability about 2.85~\mumN~\cite{Wei2017b}), and indeed, to our knowledge, is the most widely tunable pulsed fiber laser at any wavelength (e.g. exceeding 200~nm tunability at 1.9~\mum in Ref.~\cite{Meng2017}).
CW Dy:ZBLAN laser operation was recently shown between 2.8 and 3.4~\mum when pumped at 1.7~\mumN~\cite{Majewski2018}, thus even wider tunability and longer wavelength pulse generation may be possible with improved cavity designs (e.g. higher Q-factor cavities) and pump schemes.

An additional benefit of Dy is the ability to in-band pump using an Er:ZBLAN fiber laser since the laser transition is between the two lowest lying energy levels, offering Stokes-limited efficiencies exceeding 85\%.
While the 40\% slope efficiency achieved here is lower, due to excess cavity loss, this value still exceeds the theoretical maximum efficiency for Er and Ho lasers, where the 3~\mum mid-IR transition is between two excited levels.

\subsection*{Towards femtosecond mid-IR FSF lasers}
In general, the development of mode-locked lasers at novel mid-IR wavelengths is complicated by the immaturity of complementary low-loss optical components.
Wavelengths beyond $\sim$3~\mumN, for example, approach the band edge of indium-based absorber materials which are used for 2.7--2.9~\mum SESAMs~\cite{Hu2014}, forcing consideration of other less established materials.
Additionally, the lack of mid-IR compatible magneto-optical materials with a high Verdet constant raises barriers to engineering high-extinction low-loss broadband isolators, which are required for NPE-mode-locked ring cavities~\cite{Antipov2016a,Duval2015}.
Therefore, the identification of optimal mode-locking mechanisms for widely tunable mid-IR pulse generation remains an open question.

We believe that FSF using a TeO$_2$ acousto-optic crystal is a robust and versatile solution, which enabled us to construct a simple, compact pulse source with record tunability, all controlled electronically.
Self-starting mode-locking was readily achieved and remained stable for a number of hours.
Compared to previous FSF lasers with silica fiber, this is the first mid-IR FSF laser and we note that the nonlinearity of ZBLAN fiber here is an order of magnitude lower than for near-IR silica fiber (due to larger mode-fields and the reduced nonlinear index of ZBLAN); despite this, there is sufficient SPM to sustain the coherent FSF picosecond pulse generation process described earlier.
For many application, however, shorter pulse durations are required.

While the generation of tens of picosecond pulses is in agreement with the parameter space achieved from previously reported near-IR FSF lasers, we note that a number of reports have observed pulse narrowing and subsequent, sub-picosecond pulse generation through the addition of SA effects, such as including SESAMs and NPE~\cite{Kivisto2008,Noronen2016a,Heidt2007}.
We note that our cavity does include the potential to weakly exploit NPE: nonlinear polarization rotation could occur during fiber propagation and the AOTF is intrinsically polarization selective since S and P polarization components diffract at slightly different angles, and we only resonate the S component.
However, despite attempts to vary the cavity birefringence by perturbing the active fiber and explicitly adding quarter and half waveplates into the external cavity, we were unable to achieve any pulse shortening effects. 
Waveplate rotation could cause loss of mode-locking as it increased the cavity loss by rotating the cavity polarization away from the direction for which the AOTF was optimized, but no beneficial effects were observed.
This adds further support to the argument that the mode-locking mechanism in our cavity is solely driven by FSF (we note too that our GNLSE-based model is scalar, assuming only one polarization mode and the absence of such NPE effects).
As improved mid-IR isolators are developed, a ring cavity FSF laser could be adopted, where FSF facilitates simpler self-starting mode-locking than NPE alone, and the passive SA mechanism could enable shortening to femtosecond durations.

Additionally, pulse shortening in FSF lasers is possible through soliton shaping effects~\cite{Sousa2000}.
ZBLAN fiber is strongly anomalously dispersive in the mid-IR [Fig.~\ref{fig:reprate}(a)], and it is well known that even noisy waveforms in anomalous environments can converge towards a sech$^2$ soliton shape~\cite{Gouveia-Neto1989}. 
However, the lengths scales for this process can exceed the cavity length for initially broad pulses; the lack of soliton shaping in our cavity was confirmed by disabling dispersion in the simulations and observing an unchanged steady-state solution.
Early near-IR mode-locked lasers demonstrated that significant pulse shortening of 10s ps actively mode-locked systems could be achieved through cavity elongation to enhance soliton shaping~\cite{Kafka1989}, however, thus careful cavity optimization is a promising area of future work to achieve femtosecond pulses beyond 3~\mumN.

Finally, we note the FSF mechanism is ideally suited and simple to implement for other mid-IR gain materials.
As TeO$_2$ crystals are transparent up to 5~\mumN, FSF using a TeO$_2$-based AOTF would be compatible with transition metal chalcogenide lasers~\cite{Mirov2015}, $\sim$3.5~\mum Er:ZBLAN lasers~\cite{Jobin2018,Henderson-Sapir2018,Qin2018}, and even the emerging class of 4~\mum fiber laser technology based on higher-lying transitions in Ho ($\sim$3.9~\mum\cite{Maes2018}) and Dy ($\sim$4.3~\mumN~\cite{Majewski2018c}) in low-phonon-energy indium-fluoride glass.
In addition, opportunities exist to explore the impact of novel soft-glass fibers such as highly nonlinear chalcogenides on mid-IR FSF lasers, to further understand the dynamics of this novel pulse generation mechanism, which has been relatively understudied to date.

\subsection*{Applications \& future prospects}
The generation of picosecond pulses, tunable from 2.97 to 3.30~\mum ($\sim$3000--3400 cm$^{-1}$) in a flexible fiber format, creates new opportunities in the mid-IR.
This range includes a number of important functional groups (e.g. OH, NH, CH moieties), which could be targeted for sensing or processing applications.
Of particular interest is polymer machining, where enhanced ablation efficiency was recently demonstrated by exploiting resonant absorption with the CH bond compared to current 10.6~\mum CO$_2$ laser processing~\cite{Frayssinous2018}. 
Our measured picosecond pulse energy of 2.7~nJ is already approaching the record pulse energy for mid-IR mode-locked fiber lasers at shorter wavelengths (e.g. 7.6~nJ at 2.87~\mumN~\cite{Antipov2016a}), which could be power-scaled with a subsequent Dy:fiber amplifier.
Using such amplified pulses, nonlinear compression could be a simple fiber-based route for extra-cavity pulse shortening, potentially down to few-cycle durations~\cite{woodward_2017_70fs}.

Nonlinear optical applications could also exploit the laser as a pump source, e.g. for supercontinuum generation.
Pumping at longer wavelengths than currently available could enable higher efficiency broadband sources spanning the mid-IR.
Additionally, chalcogenide fibers (e.g. As$_2$S$_3$ and As$_2$Se$_3$) are often used for supercontinuum generation~\cite{Marandi2012,Hudson2017} due to their high nonlinearity, but require tapering / novel geometries to blue-shift the zero-dispersion wavelength (ZDW) closer to the pump (near-ZDW-pumping typically enables the broadest supercontinua through soliton dynamics), or alternatively, the use of interim nonlinear cascade stages~\cite{Petersen2016a}.
Thus, longer wavelength mode-locked fiber lasers could simplify supercontinuum pump requirements and relax the need for fiber dispersion engineering.

Finally, we note that the FSF concept has wider implications for optical systems development.
For example, similarities exist between the FSF mode-locking mechanism and the nascent Mamyshev Oscillator concept, whereby intensity discrimination is provided by SPM and two offset fixed-frequency filters~\cite{Liu2017a}, yet questions remain regarding optimal choice of filters, frequency offsets and the self-starting ability.
There are also opportunities for FSF frequency comb generation from coherently driven passive cavities: early FSF work highlighted this idea~\cite{Kowalski1987}, but was limited by low fiber-cavity Q factors; this could be improved by considering currently emerging high-Q micro-resonators~\cite{Pasquazi2017}.
FSF prospects are not limited to source development, however.
For example, ultrafast continuous single-shot measurements in near-IR sensing and imaging systems are being enabled by dispersive Fourier transform techniques, using long ($>$100~m) silica fiber lengths to map a pulse spectrum to a temporal waveform for measurement on a fast single-pixel detector~\cite{Goda2013}.
An equivalent mid-IR-compatible frequency-to-time mapping could be achieved through superposition of replicas of a signal shifted simultaneously in time and frequency---i.e.\ the output of a FSF cavity driven by the signal of interest~\cite{Chatelus2016}.
It is certainly an exciting prospect to reconsider FSF cavities for new mid-IR applications and further work is planned to validate such schemes.

\section{Conclusion}
In summary, we have proposed and demonstrated a new approach to mid-IR picosecond pulse generation from a compact fiber system.
The achieved spectral coverage from 2.97 to 3.30~\mum represents, to our knowledge, the most widely tunable pulsed fiber laser by a significant margin.
Within this range, stable $\sim$33~ps pulses were generated with up to 2.7~nJ energy.
This performance was achieved by leveraging the remarkable spectroscopy of dysprosium-doped fluoride fibers and the understudied frequency shifted feedback pulse generation mechanism.
Numerical simulations showed excellent agreement with our experimental measurements, permitting the exploration of radiation build-up dynamics from noise to advance understanding of the frequency shifted feedback dynamics, which is an ideal technique for spectral regions where conventional mode-locking techniques (e.g. saturable absorbers) are unavailable / lack broadband tunability.
The long wavelength and broad spectral coverage of mode-locked dysprosium fiber lasers thus opens up new opportunities in the mid-IR.

\section*{Acknowledgment}
This work was funded by the Australian Research Council (ARC) (DP140101336, DP170100531).
RIW also acknowledges support through an MQ Research Fellowship.



\end{document}